\documentclass[11pt,a4paper]{article}
\usepackage{graphicx}

\newcommand{\myblstr}{1}  
\renewcommand{\baselinestretch}{\myblstr}

\newcommand{\be}{\begin{equation}}
\newcommand{\ee}{\end{equation}}
\newcommand{\bea}{\begin{eqnarray}}
\newcommand{\eea}{\end{eqnarray}}
\newcommand{\bref}[1]{(\ref{#1})}
\renewcommand{\th}{\theta}
\newcommand{\eps}{\epsilon}
\newcommand{\pd}[1]{\partial_{{#1}}}
\newcommand{\dr}{\partial_r}
\newcommand{\dth}{\partial_\th}
\newcommand{\Lag}{{\cal L}}
\newcommand{\Lagf}{{\cal L}_{\rm fermions}}
\newcommand{\hconj}{(\mbox{h. c.})}
\newcommand{\sig}[1]{\sigma^{#1}}
\newcommand{\mass}[1]{m_{\rm #1}}
\newcommand{\mgut}{m_{\rm \scriptscriptstyle GUT}}
\newcommand{\mup}{\mass{u}}
\newcommand{\mL}{m_{\! L}}
\newcommand{\mR}{m_{\! R}}
\newcommand{\ph}[1]{\Phi_{#1}}
\newcommand{\pharrow}[1]{\stackrel
	{ \rule[-2pt]{0.0 in}{2 pt} \scriptstyle \ph{{#1}} }{\longrightarrow}}

\newcommand{\spinU}{\mbox{\scriptsize 
		$\left(\begin{array}{@{}c@{}}1\\0\end{array}\right)$}}

\newcommand{\bmat}[1]{\left(\begin{array}{{#1}}}
\newcommand{\emat}{\end{array}\right)}

\begin{document}

\title{\bf Destruction of Fermion Zero Modes \\ on Cosmic Strings}

\author{Stephen C. Davis\thanks{
Department of Physics, University of Wales Swansea,
Singleton Park, Swansea, SA2 8PP, Wales.}}

\maketitle

\begin{abstract}

I examine the existence of zero energy fermion solutions (zero modes)
on cosmic strings in an $SO(10)$ grand unified theory. The current
carrying capability of a cosmic string formed at one phase transition
can be modified at subsequent phase transitions. I show that the zero
modes may be destroyed and the conductivity of the string altered. I
discuss the cosmological implications of this, and show that it allows
vorton bounds to be relaxed. 

\end{abstract}

\section{Introduction}

Cosmic strings, which are a type of topological defect, arise in many
grand unified theories. Large quantities of them may be produced at
phase transitions in the early universe. A network of cosmic strings
could explain the observed anisotropy in the microwave background
radiation and the large scale structure of the
universe~\cite{string book}. 

It has been realised in the past few years that cosmic strings have a
far richer microstructure then previously thought~\cite{wbp&acd}. In
particular, the presence of conserved currents in the spectrum of a
cosmic string has profound implications for the cosmology of the
defects. In this paper I examine currents with fermion
charge carriers, particularly massless ones. 

Currents may provide a method of detecting strings. If they form at
high energy scales, and are charged, the resulting electromagnetic
field may be detectable~\cite{Witten}. Decaying currents may explain
observed high-energy cosmic rays~\cite{Witten,current-rays}, and could
also provide a mechanism for baryogenesis~\cite{acd&wbp3,currbaryo}. There are
several processes which can create currents on strings. These include
interaction with the plasma and collisions between cosmic
strings. Charged currents can also be generated by magnetic or
electric fields. Stable relics, vortons, can form as collapsing string
loops are stabilized by the angular momentum of the trapped charge
carriers~\cite{vorton}. If they form at high energy scales, these
relics can dramatically alter the evolution of string
networks. Vortons formed at low energies may provide a dark matter
candidate~\cite{string book}. 

As the universe expands, a network of cosmic strings will stretch,
suggesting that its energy density will grow relative to everything
else. This would lead to the universe becoming string-dominated, which
is strongly ruled out by observation. More careful analysis shows that
loops of string will break off from the network. These then contract,
losing energy by gravitational radiation. This avoids string
domination. If the loops are stabilized by conserved currents, this
mechanism fails, allowing the corresponding theory to be severely
constrained~\cite{vortonbounds}. 

The cosmological implications of the vortons are most pronounced when
the universe has become matter-dominated. If they decay during the era
of radiation domination, the cosmological catastrophe may be avoided
and the vorton bounds evaded. It has recently been realized that
subsequent phase transitions can have a considerable effect on the
microphysics of cosmic strings. Massless fermion currents on cosmic
strings can be both created \cite{acd&wbp} and destroyed
\cite{SO10strings}. In this paper I examine the destruction of such
currents in grand unified theories with an $SO(10)$ symmetry group. 

I describe a simple cosmic string model in Section~\ref{sec ZeroModes}, 
and show that it has zero energy fermion solutions, called
zero modes. This result was originally derived by Jackiw and
Rossi~\cite{Jackiw}. I then show how the zero-mode solutions can be
extended to massless currents. 

In Section~\ref{sec SO10} I discuss the fate of zero modes on strings
formed in an $SO(10)$ grand unified symmetry~\cite{SO10strings}. At
high temperatures the theory resembles the toy model of
Section~\ref{sec ZeroModes}. I investigate the implications of the
electroweak phase transition for the zero modes, and show that they do
not survive it. 

I consider the implications of spectral flow in
Section~\ref{sec Spectral}, and deduce that the currents acquire a
small mass. The string current can now dissipate. This allows vortons
to decay and weakens the cosmological bounds on such models
\cite{acd&wbp3}. An important feature that allows zero modes to be
removed is the presence of a massless particle that mixes with the
zero mode after the transition. Finally, I summarize the conclusions.

\section{Fermion Zero Modes on an Abelian Cosmic String}
\label{sec ZeroModes}

The Abelian Higgs model has the Lagrangian
\be
\Lag = (D_\mu \phi)^\ast(D^\mu \phi) 
		- \frac{1}{4}F_{\mu \nu} F^{\mu \nu} - 
\frac{\lambda}{4} (|\phi|^2 - \eta^2)^2
\label{U1model}
\ee
with $F_{\mu \nu} = \pd{\mu} A_\nu - \pd{\nu} A_\mu$. As well
as the usual $\phi=$ const.\ solution, it has cosmic string
solutions of the form
\bea
\phi &=& \eta f(r) e^{in\th} \label{U1Ansatza} \\
A_\mu &=& n\frac{a(r)}{er}\delta^\th_\mu
\label{U1Ansatzb}
\eea
In order for the solution to be regular at the origin
$f(0)=a(0)=0$. If the solution is to have finite energy, $f(r)$ and
$a(r)$ must tend to 1 as $r \rightarrow \infty$. The resulting string
is the well known Nielsen-Olesen vortex~\cite{Nielsen}.  It turns out
that $f$ and $a$ take their asymptotic values everywhere outside of a
small region around the string. Thus $|\phi|$ is constant and $A_\mu$
is pure gauge away from the string. The size of this region is of
order $\eta^{-1}$. 

Consider an extension of \bref{U1model} to include a two-component
fermion, with charge $1/2$. The extra terms in the Lagrangian will then be
\be
\Lagf = \bar{\psi} i\sig{\mu} D_\mu \psi 
- \frac{1}{2} \left[ig_Y \bar{\psi} \phi \psi^c + \hconj \right] 
\label{Lagferm}
\ee
where $\sig{\mu} = (-I,\sig{i})$, $D_\mu\psi = \left(\pd{\mu} -
\frac{1}{2}ieA_\mu\right)\psi$, and $\psi^c = i\sig{2} \psi^\ast$ 
is the charge conjugate of $\psi$. This gives the field equations
\be
\bmat{cc} 
-e^{i\th}\left[\dr +\frac{i}{r}\dth +n\frac{a(r)}{2r} \right] &
\pd{z} + \pd{t} \\ \pd{z} - \pd{t} & 
e^{-i\th}\left[\dr -\frac{i}{r}\dth -n\frac{a(r)}{2r} \right] 
\emat \psi  - \mass{f} f(r) e^{in\th} \psi^\ast = 0
\label{U1Ferm} \ee
where the expressions \bref{U1Ansatza} and \bref{U1Ansatzb} have
been substituted for $\phi$ and $A_\mu$, and $\mass{f}=g_Y \eta$. 

Following previous work by Jackiw and Rossi~\cite{Jackiw} we look
for solutions with no $z$ or $t$ dependence. Such solutions will have
zero energy and are called zero modes. We can see from \bref{U1Ferm}
that they can also be taken as eigenstates of $\sig{3}$. The angular
dependence of a solution satisfying $\sig{3}\psi = \psi$ can be
separated out with the ansatz
\be
\psi(r,\th) = \spinU \left(U(r) e^{il\th} + V^\ast(r)e^{i(n-1-l)\th}\right)
\label{U1ansatz}
\ee
$l$ is an arbitrary integer. We will consider the case $2l=n-1$
separately. The field equations \bref{U1Ferm} imply
\bea
\left(\pd{r} - \frac{l}{r} + \frac{na(r)}{2r}\right)U + \mass{f} f V &=& 0
\nonumber \\
\left(\pd{r} - \frac{n-1+l}{r} + \frac{na(r)}{2r}\right)V 
+ \mass{f} f U &=& 0
\label{U1equ}
\eea
Analytic solutions to these equations cannot generally be
found. However it is possible to determine the number of physical
solutions by considering their asymptotic form. 

At large $r$ we can approximate $f$ and $a$ by 1. Then \bref{U1equ} can
be solved with modified Bessel functions, which are
asymptotically equal to $e^{\pm \mass{f} r}/\sqrt{r}$. We are
interested in normalizable solutions, with $\int |\psi|^2 \, d^2 x$
finite, so only the decaying solution is acceptable. 

When $r$ is small, we can neglect $f$ and $a$. To first order the
solutions of \bref{U1equ} are determined by the angular dependence of
\bref{U1ansatz}
\bea 
U &\sim& r^l \ , \ \ V = O(r^{l+1}) \nonumber \\
V &\sim& r^{n-1-l} \ , \ \ U = O(r^{n-l})
\eea

In order to match up with the one acceptable large-$r$ solution, both
the small-$r$ solutions must be regular. This will only be true if $0
\leq l \leq n-1$. This suggests there are $n$ solutions of
\bref{U1equ} if $n>0$ since there are $n$ choices of $l$. In fact we
have counted each solution twice, since putting $l \rightarrow n-1-l$
in \bref{U1ansatz} gives an equivalent ansatz. For every real solution
of \bref{U1equ} there is also an imaginary one. This gives a total of
$n$ real solutions.

The above analysis needs modification when $l=n-1-l$. We can set
$V^\ast = U$ in \bref{U1ansatz}, giving the single equation
\be
\left(\pd{r} - \frac{l}{r} + \frac{na(r)}{2r}\right)U 
+ \mass{f} f U^\ast = 0
\ee
This can be solved analytically. The solution which is well-behaved at
large $r$ is 
\be
\psi(r,\th) = \spinU r^l \exp\left(
-\int^r_0 \mass{f} f(s) + n\frac{a(s)}{2s} ds\right)
\label{aU1FermSol}
\ee
This is regular at $r=0$ if $l=(n-1)/2 \geq 0$.

Similar analysis can be applied to the solutions satisfying
$\sig{3}\psi = -\psi$. We find a total of $|n|$ solutions, which are
all eigenstates of $\sig{3}$. Their eigenvalues are $+1$ if $n>0$, and
$-1$ if $n<0$. All the solutions decay exponentially outside the
string, and so are confined to it. They can be regarded as fermions
trapped on the string. 

We can see from \bref{U1Ferm} that the solutions can easily be
extended to include $z$ and $t$ dependence. This is achieved by
multiplying $\psi$ by $\alpha(z,t)$, which satisfies $(\pd{z} \mp
\pd{t})\alpha = 0$, depending on whether $\sig{3}\psi=\pm\psi$. Thus
the trapped fermions move at the speed of light in the $\pm z$
direction. 

These currents are conserved, and the string acts as a perfect
conductor. While zero modes have very little cosmological
significance, the lightlike currents can have dramatic consequences. 

\section{An $SO(10)$ GUT with Strings}
\label{sec SO10}

One example of a phenomenologically credible grand unified theory (GUT)
has the symmetry breaking
\bea
SO(10) & \pharrow{126} & SU(5) \times Z_2 \nonumber \\ 
& \pharrow{45} & SU(3)_c \times SU(2)_L \times U(1)_Y \times Z_2 \nonumber \\
& \pharrow{10} & SU(3)_c \times U(1)_Q \times Z_2
\eea
The discrete $Z_2$ symmetry allows the formation of topologically
stable cosmic strings. One possibility is an Abelian string, similar to that 
described in Section~\ref{sec ZeroModes},
\bea 
\ph{126} &=& e^{in\th} \phi_0 f(r) \\
X_\th &=& \frac{n}{\sqrt{10}}\frac{a(r)}{er}
\eea
$\phi_0$ is the usual VEV of $\ph{126}$, and $X$ is a GUT gauge field
broken by $\ph{126}$. The only stable Abelian strings have $|n|=1$,
but higher winding number strings may have long lifetimes. Non-Abelian
strings also form in this model, but they do not have zero modes at
high temperatures, and I will not consider them
here~\cite{SO10strings}.

The string gauge field has a nontrivial effect on the electroweak
Higgs field $\ph{10}$. The components of $\ph{10}$ have charges $\pm
1/5$ with respect to the generator of the GUT gauge field $X$. If $n$
is high enough, it is energetically favourable for $\ph{10}$ to wind
like a string.
\bea
\ph{10} &=& H_u e^{im\th} h_u (r)+ H_d e^{-im\th} h_d (r) \\
Z_\th &=& \sqrt{\frac{5}{8}}\left(m-\frac{n}{5}\right)\frac{b(r)}{er}
\eea
where $H_u$ and $H_d$ are the usual VEVs of the components of
$\ph{10}$. Whether $\ph{10}$ winds or not, a nonzero electroweak gauge
field is required to give a vanishing covariant derivative outside the
string. The solution satisfies the boundary conditions
$h_u(\infty)=h_d(\infty)=b(\infty)=1$, $b(0)=0$ and $h_u(0)=h_d(0)=0$
if $m \neq 0$. If $m=0$ $h_u$ and $h_d$  are small, of order
$|H_{u,d}|/|\phi_0|$~\cite{SO10strings}.  $m$ is determined by the GUT
string, and is equal to the nearest integer to $n/5$, so the
electroweak Higgs field does not wind around a topologically stable
Abelian string. The region of electroweak symmetry restoration is
inversely proportional to the electroweak scale, so is far greater
than the string core. 

The fermion sector of the $SO(10)$ GUT contains all the Standard Model
fermions, and an extra right-handed neutrino for each family. For
simplicity I will just consider the string's effect on just one family,
although it is easy to generalize the results. Of the three Higgs
fields, only $\ph{126}$ and $\ph{10}$ can couple to fermions. Only
right-handed neutrinos couple to $\phi_0$, while neutrinos of either
helicity couple to $H_u$. As the gauge symmetry unifies left- and 
right-handed particles, it is convenient to express everything in terms of
left-handed spinors. I will use $\nu = \nu_L$ and $\nu^c
=i\sigma^2\bar{\nu}_R^T$ to express neutrino terms. Defining
$\psi^{(\nu)} = (\nu^c,\nu)^T$, the resulting neutrino mass terms are
\be
\bar{\psi}^{(\nu)}\bmat{cc} \mgut f(r)e^{in\th} & \mup h_u(r) e^{im\th} \\ 
	\mup h_u(r) e^{im\th} & 0 \emat \psi^{(\nu)}
\ee
$\mup \sim |H_u| \sim 1$ MeV is the up-quark mass and
$\mgut \sim |\phi_0| \sim 10^{16}$ GeV. 
Since $\eps = \mup / \mgut \ll 1$, the neutrino mass eigenvalues
outside the string are
\bea
\mR &=& \mgut \frac{\sqrt{1 + 4\eps^2} + 1}{2} 
 \approx \mgut \nonumber \\
\mL &=& \mgut \frac{\sqrt{1 + 4\eps^2} - 1}{2} 
 \approx \frac{\mup^2}{\mgut}
\eea

The mass eigenstates are then approximately $\nu^c + \eps \nu$ and
$\nu - \eps \nu^c$. This illustrates the
seesaw mechanism~\cite{seesaw}. Although the neutrinos have the same
couplings to the electroweak Higgs field as the up quark, the GUT
Higgs ensures that $\nu_R$ is superheavy and $\nu_L$ is very light, as
is required to agree with observation. Recent measurements have suggested
that $\nu_L$ does indeed have a small mass~\cite{numass}.

At high temperatures only $\ph{126}$ is nonzero and $\mup=0$. The
model then reduces to the toy model discussed in
Section~\ref{sec ZeroModes}, with $\psi=\nu^c$. Thus $|n|$ right-handed
neutrino zero modes exist on the string. This implies that such strings
always have conserved currents at high temperatures. These could allow
vortons to be formed. If string loops formed at this energy scale do
not decay, their evolution will lead to serious conflict with
observations of the present universe~\cite{vortonbounds}. 

The situation is more complex for the neutrino fields at low
temperatures since they couple to two Higgs fields at the same
time. As in Section~\ref{sec ZeroModes}, we will first look for
solutions satisfying $\sig{3}\psi=\psi$. With respect to the generator
of the $X$ gauge field, $\nu$ has charge $3/10$. It has the same
charge as $\ph{10}$ with respect to the $Z$ generator.
The resulting field equations are
\be
e^{i\th}\left(\dr + \frac{i}{r}\dth + \frac{n}{2}\frac{a(r)}{r} \right)
\nu^c + \mup h_u(r) e^{im \th} \nu^\ast
+ \mgut f(r) e^{in\th} \nu^{c\ast} = 0 \label{abferNeuc}
\ee \be
e^{i\th}\left(\dr + \frac{i}{r}\dth  - \frac{3n}{10}\frac{a(r)}{r} 
+ \left[m-\frac{n}{5}\right]\frac{b(r)}{r}  \right)
\nu + \mup h_u(r) e^{im\th} \nu^{c\ast} = 0 \label{abferNeu}
\ee
Although Jackiw and Rossi did not consider this
case, it can be approached using a similar method to theirs.
The angular dependence can be removed with the substitutions
\be
\nu^c = \spinU \left(U e^{il\th} + V^\ast e^{i(n-1-l)\th}\right)
\label{abferNeuThc}
\ee \be
\nu = \spinU \left(W^\ast e^{-i(m+1+l)\th} + Y e^{-i(m+n-l)\th}\right)
\label{abferNeuTh}
\ee
(The case $2l = n-1$ will be considered later.) The resulting four
complex differential equations are
\be
\left(\dr - \frac{l}{r} + \frac{na(r)}{2r} \right) U 
+ \mup h_u(r) W + \mgut f(r) V = 0 
\label{abNeua} \ee \be
\left(\dr - \frac{n-1-l}{r} + \frac{na(r)}{2r} \right) V 
+ \mup h_u(r) Y + \mgut f(r) U = 0
\label{abNeub} \ee \be
\left(\dr + \frac{m+1+l}{r} + \frac{(10m - 2n)b(r)-3na(r)}{10r}\right) W 
+ \mup h_u(r) U = 0
\label{abNeuc} \ee \be
\left(\dr + \frac{m+n-l}{r} + \frac{(10m - 2n)b(r)-3na(r)}{10r}\right) Y
+ \mup h_u(r) V = 0
\label{abNeud}
\ee
Splitting these equations into real and imaginary parts gives two
identical sets of four real equations, and so it is only necessary to look
for real solutions.

When $r$ is large, $h_u$, $f$, $a$, and $b$ are all approximately 1. As 
with \bref{U1equ}, rearrangement of
\bref{abNeua}--\bref{abNeud} reduces them to modified Bessel equations at
large $r$. The four independent solutions are proportional to $e^{\pm
\mR r}/\sqrt{r}$ and $e^{\pm \mL r}/\sqrt{r}$, so only two of them are
normalizable. 

For small $r$, $h_u \sim r^{|m|}$, $f \sim r^{|n|}$, and the gauge
terms are of order $r$. Thus we can ignore them when finding the
leading-order terms of the small-$r$ solutions of
\bref{abNeua}--\bref{abNeud}. The dominant terms of the four independent
solutions are
\be
r^l \ , \ \ \ r^{n-1-l} \ , \ \ \ 
r^{-m-1-l} \ , \ \ \ r^{l-n-m}
\label{abNeuZeroSol}
\ee

If $\varphi$ is a solution of \bref{abNeua}--\bref{abNeud} for all $r$,
which is normalisable at $r=\infty$, then it will have to match some
combination of the two normalizable solutions for large $r$.  At $r=0$,
$\varphi$ will be made up of a combination of the solutions in
(\ref{abNeuZeroSol}). So if $\varphi$ is to be normalizable
everywhere, at least three of the solutions \bref{abNeuZeroSol} must be
well behaved at $r=0$.  Thus for each $l$ satisfying three of the
inequalities $l \geq 0$, $l \leq -m-1$,  $l \leq n-1$ and $l \geq
n+m$, there will be one normalizable solution. If $l$ satisfies all four
there will be two solutions. Not all of these solutions are independent
since the real (or imaginary) solutions for $l=l'$ and $l=n-1-l'$ are
proportional. 

For $l = (n-1)/2$ the angular dependence of \bref{abferNeuc}
and \bref{abferNeu} is removed with the substitutions
\be
\nu^{c} = U e^{il\th} \ , \ \ \ \ \nu = W^\ast e^{-i(m+1+l)\th}
\ee
giving (after dropping gauge terms) the equations 
\be
\left(\dr - \frac{l}{r} + \frac{na(r)}{2r}\right) U
+ \mup h_u(r) W + \mgut f(r)U^{\ast} = 0
\ee \be
\left(\dr + \frac{m+1+l}{r} + \frac{(10m - 2n)b(r)-3na(r)}{10r}\right) W 
+ \mup h_u(r) U = 0
\ee

The two real solutions have the asymptotic forms $e^{-\mR r}/\sqrt{r}$
and $e^{\mL r}/\sqrt{r}$, while the two imaginary solutions are
$e^{\mR r}/\sqrt{r}$ and $e^{-\mL r}/\sqrt{r}$. The leading-order
terms of the small-$r$ solutions (real or imaginary) are
\be
r^l \ , \ \ r^{-m-1-l}
\ee
Matching large- and small-$r$ solutions reveals that in this case there
is one real and one imaginary solution if  $0 \leq l \leq -m-1$. 

Combining all the above results gives a grand total of $2m$ ($m$ real
and $m$ imaginary) normalizable solutions if $m > 0$, and $0$
otherwise. Surprisingly, this does not depend on $n$. A similar
approach can be applied to the other components of $\nu^c$ and $\nu$
to give $-2m$ normalizable solutions, provided $m<0$. Hence there are
$2|m|$ possible neutrino zero modes after electroweak symmetry
breaking. Examination of the asymptotic zero-mode solutions reveals
that they are confined to the region of electroweak symmetry
restoration. 

Since $|2m| < |n|$, some of the zero modes will be destroyed. For a
stable $n=1$ string all zero modes are destroyed.  Thus, since higher $n$
strings almost certainly decay, there are zero modes before, but not
after the electroweak phase transition. The neutral current in the
string disperses~\cite{HillWidrow} and any vortons formed would
dissipate after about $10^{-10}$ sec~\cite{acd&wbp3}. Before the electroweak
phase transition from about $10^{10}$ GeV to $10^2$ GeV the universe
would undergo a period of matter domination. Once the vortons
dissipate there would be some reheating of the universe. However, the
electroweak interactions and physics below the phase transition would
be unaffected.

Although we have looked at a specific $SO(10)$ symmetry breaking, the
results apply to most other breakings of this group, since they have
the same fermion mass terms. In a more arbitrary
$U(1) \times $(Standard Model) theory, the ratio of $m$ and $n$ could
be greater than 1, in which case extra zero modes would be created at the
electroweak phase transition. 

Although we have only considered one type of model, the arguments
can be applied to any theory with fermions coupling to
cosmic strings. This has been done for a general theory in
ref.~\cite{Index}. The results obtained agree with simpler index
theorems derived previously~\cite{WeinbergGanoulis}. 

\section{Index Theorems and Spectral Flow}
\label{sec Spectral}
\renewcommand{\baselinestretch}{1}

\begin{figure}
\vskip 0.5 in
\centerline{\includegraphics[width=3.5in]{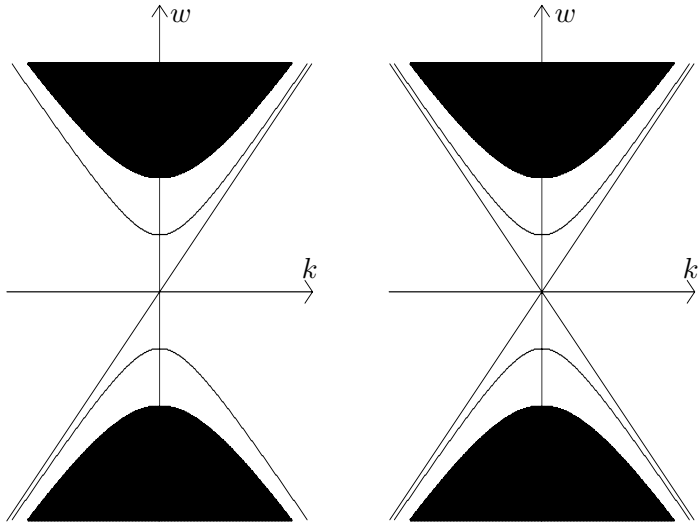}\ \hspace{0.6in} \ }
\vskip -3.2 in 
\hskip 2.02 in $w$
\hskip 1.85 in $w$
\vskip 1.15 in 
\hskip 2.71 in $k$
\hskip 1.86 in $k$
\vskip 1.2 in
\caption{The Dirac spectrum with a zero mode (left) and a very low
lying bound state (right). Both spectra also have a bound state and
continuum.}
\label{specfig}
\end{figure}
\renewcommand{\baselinestretch}{\myblstr}

I have shown that zero modes can acquire masses at subsequent phase
transitions. No matter how small this mass, the spectrum of the Dirac
operator changes significantly. If we compare the Dirac spectrum with
a zero mode and a low-lying bound state with infinitesimal mass
(Fig.~\ref{specfig}), we see that an arbitrarily small perturbation to
the zero mode introduces an entire new branch to the spectrum. Any
massive state gives a spectrum that is symmetric about both the $w$
and $k$ axes; there is always a reference frame in which the particle
is at rest and others where it is moving up or down the
string. Conversely, the zero mode, which is massless, can only move in
one direction along the string and its spectrum is asymmetric.  The
transition from zero mode to low-lying bound state causes drastic
changes in the spectrum and can be brought about by infinitesimal
changes in the value of one Higgs field. If we consider the species
with the zero mode alone, this infinite susceptibility to the
background fields appears unphysical. However, when we include the
massless neutrino in the $SO(10)$ model the spectral changes are less
worrying. For a small coupling between the two neutrinos, both the
before and after spectra have a continuum of massless or nearly
massless states. These states can be used to build the extra branch of
the perturbed zero-mode spectrum, allowing small changes in the
overall spectrum for small changes in the background fields. This
observation leads me to conjecture that zero modes can be removed only
if they become mixed with other massless states. 

At the electroweak phase transition the neutrino zero mode will mix
with the left-handed neutrino field to form a bound state. Its mass
will be proportional to the neutrino mass inside the string, which is
of order $\mup^3/\mgut^2$~\cite{bound}. High energy currents are free
to scatter into left-handed neutrinos off the string, so the maximal
current will be very small. This will be insufficient to stabilize
vortons. Additionally, the current will be spread over the region of
electroweak symmetry restoration. This is far larger than the size of
a vorton, which is a couple of orders of magnitude greater than the
GUT string radius~\cite{vorton}. The current on one part of the string
will then interact with current on the opposite side of the loop,
increasing the vorton's instability. Thus vortons in $SO(10)$ will certainly
decay at the electroweak phase transition. 

Although I have discussed a specific type of GUT, many of the ideas
I have used apply to a far wider range of theories. Fermion zero
modes are generic in supersymmetric theories~\cite{SUSY}. They, too, can
be destroyed at lower temperatures, in this case by supersymmetry
breaking~\cite{SUSY2}. 

\section{Conclusions}

If fermions couple to a cosmic string Higgs field, their spectrum will
gain extra states which are confined to the string core. The existence
of zero-energy fermion states on strings can be examined
analytically. Such solutions can easily be extended to give massless
currents, which will have significant cosmological effects. In this
paper I have investigated the existence of fermion zero modes in two
simple models, one of which is contained in a realistic $SO(10)$
GUT. We have seen that the microphysics of cosmic strings can be
influenced by subsequent phase transitions. Fermion zero modes, and
consequently lightlike currents on the strings can be created or
destroyed by such phase transitions. In determining whether or not a
cosmic string carries conserved currents it is not enough to just
consider them at formation, but one must follow the microphysics through the
multiple phase transitions that the system undergoes.

It is possible
for vortons formed at high energy to dissipate after a subsequent
phase transition if the relevant fermion zero mode does not survive
the phase transition, thus vorton bounds could be evaded. I have
demonstrated this effect by analysing the neutrino zero modes of an
$SO(10)$ model in detail. Prior to dissipation there could be a period
of vorton domination; after the phase transition the universe would
reheat and then evolve as normal. 

The right-handed neutrino zero modes are removed at the electroweak
phase transition when they mix with the Standard Model left-handed
neutrinos. By considering spectral flow we have seen that zero modes and
massless currents can only be removed by mixing them with another
massless field. The resulting state will be a bound state or massive
current. In the model I considered, the maximal current that the
string can support after the electroweak phase transition is
proportional to the left-handed neutrino mass. Such currents are far
too small to stabilize vortons. 

\section*{Acknowledgements}

I wish to thank my collaborators Anne C. Davis and Warren B. Perkins
for all their help. I also wish to thank Trinity College, Cambridge
and the Universtity of Wales Swansea for financial support. I am
especially grateful to the organisers of the third Peyresq meeting on
cosmology for providing such an enjoyable and interesting conference, and
for inviting me.

\end{document}